\documentclass[article,12pt]{article}
\usepackage{authblk}
\usepackage{mathrsfs}

\makeatletter %
  \renewcommand{\@biblabel}[1]{#1.}
\makeatother

\usepackage{natbib}
\usepackage{overpic}
\usepackage{color}

\usepackage{geometry}
\usepackage{amssymb}
\usepackage{lineno}
\title{Circling in on convective organization}

\author[1]{Jan O. Haerter}
\author[2]{Steven J. B\"oing}
\author[1]{Olga Henneberg}
\author[1]{Silas Boye Nissen}

\affil[1]{Niels Bohr Institute, University of Copenhagen, Blegdamsvej 17, 2100 Copenhagen, Denmark}
\affil[2]{School of Earth and Environment, University of Leeds, Leeds, LS2~9JT, UK}

\begin{document}

\maketitle
\noindent
\section*{Corresponding author}
Jan O. Haerter (haerter@nbi.ku.dk)\\
Silas Boye Nissen (silas@nbi.ku.dk)

\section*{Key Points}
\begin{itemize}
 \item Convection is often initiated by the collision of three cold pools. 
 \item A model based on expanding and colliding circles captures convective scale increase. 
 \item This simple model can produce clustering of precipitation cells. 
\end{itemize}

\pagebreak

\small
\noindent
\section*{Abstract} 
Cold pools (CPs) contribute to convective organization. However, it is unclear by which mechanisms organization occurs. 
By using a particle method to track CP gust fronts in large eddy simulations, we characterize the basic collision modes between CPs.
Our results show that CP interactions, where three expanding gust fronts force an updraft, are key at triggering new convection.
Using this, we conceptualize CP dynamics into a parameter-free mathematical model:
circles expand from initially random points in space. 
Where two expanding circles collide, a stationary front is formed. 
However, where three expanding circles enclose a single point, a new expanding circle is seeded.
This simple model supports three fundamental features of CP dynamics: 
precipitation cells constitute a spatially interacting system;
CPs come in generations;
and scales steadily increase throughout the diurnal cycle. 
Finally, this model provides a framework for how CPs act to cause convective self-organization, clustering, and extremes.

\section*{Plain language summary}
Cold pool (CP) dynamics constitutes a crucial organizing mechanism for mid-latitude and tropical clouds --- they play a key role in the lead-up to extreme events, and may influence how such events behave in a changing climate.  
CPs are dense air masses that form under precipitating thunderstorm clouds.
Under gravity, CPs spread along the surface and stimulate new precipitation events when they collide with other CPs. 
We show that CP interaction can be captured by a simple model, where circles grow in space and form new circles when three of them collide. 
Generalizing to thousands of initial circle centers, the dynamics of these circles gives a steady scale increase over time, similar to the one found in high-resolution atmospheric simulations.
In summary, we introduce a cloud organizing mechanism that forms the basis for extreme convective precipitation events, such as those implicated in flash floods.


\section{Introduction}
The transition from a non-precipitating to a precipitating atmosphere fundamentally alters the organization of its cloud field \citep{tompkins2001organizationCold,Feingold2010,Koren2011,moseley2016,haerter2017precipitation,zuidema2017survey}. 
Cold pools (CPs) are produced when rain evaporates or other hydrometeors melt, forming boundary layer air that is denser than the surrounding air. 
Due to its larger density and gravity CP air accelerates towards the surface and is then forced to spread laterally. 
From large eddy simulations (LES) and from observations it is well-established that such CP outflows act to organize the sub-cloud temperature, moisture and wind fields \citep{purdom1982subjective,droegemeier1985three,tompkins2001organizationCold,khairoutdinov2006high,feng2015mechanisms,schlemmer2015modifications,torri2016rain,de2017cold}.
Apart from influencing the thermodynamic stability in the boundary layer \citep{tompkins2001organizationCold,torri2015mechanisms}, by which CP edges may become relatively buoyant, CPs play a crucial role as a dynamic trigger for convection \citep{jeevanjee2015effective,torri2015mechanisms,moseley2016,haerter2018intensified}.
Moreover, studies have shown that in particular collisions between CPs are collocated with strong updrafts \citep{lima2008convective,boing2012influence,feng2015mechanisms}.
CP fronts caused by deep convection can travel at velocities on the order of $10\;m\;s^{-1}$ radially in tropical environments (CPs can move faster for midlatitude storms) \citep{romps2016sizes,drager2017characterizing,zuidema2017survey}. 
When CPs collide, momentum conservation can displace moist air vertically \citep{torri2015mechanisms,moseley2016,cafaro2018characteristics}.
This air thereby can condense, overcome the level of free convection and rise further due to increased buoyancy.

To better represent the effects of CPs on convection, and obtain a more realistic timing of the diurnal cycle of precipitation, some general circulation models include a statistical representation of CP processes, 
e.g. by describing the population of CPs as circular objects of identical radii \citep{rio2009shifting,grandpeix2010density,grandpeix2010densityA}.
Furthermore, extreme convective precipitation increases at rates beyond that of the saturation mixing ratio, as expressed through the Clausius-Clapeyron relation \citep{Lenderink:2008,berg2013strong,haerter2018intensified,guerreiro2018detection}. 
Explanations may well require arguments involving CP-induced self-organization \citep{moseley2016,haerter2018intensified}.

CP dynamics has been described by simplified models \citep{bengtsson2013stochastic,boing2016object}, showing that they can act as two-dimensional cellular automata. 
However, the self-organization of convection is still poorly understood, especially how small-scale interaction may give rise to large-scale organization \citep{moncrieff2010multiscale}. 
In previous simulations of the diurnal cycle of convection \citep{haerter2017precipitation}, it was found using a spatial correlation function, that the velocity field is initially characterized by Rayleigh-B\'enard-type convection during the morning hours, with lateral scales set by the boundary layer height \citep{mellado2012direct}.
As soon as precipitation sets in near mid-day, this correlation function shows a systematic increase of the distance between updrafts, indicating that the onset of precipitation leads to an increase of typical scales. 
Comparing various simulations suggested that the scale increases approximately linearly with time and the rate of increase is not strongly dependent on the simulation boundary conditions.

Our aim is to isolate the key interactions between CPs from LES to incorporate these in a simple conceptual model, and to analyze how they influence the cloud field.

\section{Data and Methods}\label{sec:methodology}
\subsection{Large eddy simulation data}\label{sec:data} 
We use simulation data from the University of California, Los Angeles, Large Eddy Simulation code (UCLA-LES \citep{stevens2005}) for the diurnal cycle of convection under spatially homogeneous surface boundary conditions and a prescribed, temporally varying surface temperature cycle.
The model domain size is $192\;km\times 192\;km$ horizontally and $17\;km$ vertically. 
The simulation is run at $200$ $m$ horizontal grid spacing and has $75$ vertical model levels stretching from $100$ $m$ at the surface to $400$ $m$ at the domain top.
The model uses a two-moment microphysics scheme \citep{seifert2006two} and a delta four-stream radiation code \citep{pincus2009monte}. 
Surface latent and sensible heat fluxes are modeled using Monin-Obukhov similarity theory.
The model time step ranges from $1$ to $5$ $s$.
The atmosphere is initialized by vertical profiles for temperature and humidity observed at Lindenberg, Germany, during the summers of 2007 and 2008 (Fig. S3 of \citet{haerter2018intensified}).
No ambient wind or pressure gradients are applied.
For the results in Figs~\ref{fig:vertical_velocity} and S4, 
the surface temperature $T_s(t)$ varies sinusoidally with an amplitude of $10$ $K$, a period of $24\;h$, a peak at $12\;h$ (noon), and a mean $\overline{T}_s=27$ $^{\circ}C$.
The comparison of scale increases in Fig.~\ref{fig:results}a makes use of data from six large eddy simulations with distinct boundary conditions.
The symbols in Fig.~\ref{fig:results}a refer to:
$\overline{T}_s=23^{\circ}C$ (CTR);  $24^{\circ}C$ (P1K);  $25^{\circ}C$ (P2K);  $27^{\circ}C$ (P4K, identical to the one discussed above); as well as a simulation where the period of forcing was increased to $48$ $h$ (LD) but otherwise maintaining CTR; and finally a less stable initial atmospheric lapse rate (LAPSE).
These simulations and the data plotted in Fig.~\ref{fig:results}a are described in more detail in \citet{haerter2017precipitation} and \citet{moseley2016}. 

\subsection{Particle method for CP tracking}\label{sec:particle_method}
We here employ a method for tracking of CPs, which capitalizes on the observation that forced uplift is crucial in creating the sharp updrafts required for new cell initiation.
Our method is described as follows:

\noindent
{\bf Defining rain events.}\\
A {\it rain cell} is defined as a spatially contiguous patch formed by grid boxes which each exceed an accumulated precipitation intensity limit $I_0$ (set to $I_0=0.5\;mm\;h^{-1}$) during an interval $\Delta t$, with $\Delta t=5\;min$. 
A lower limit of $10$ grid boxes is required for a patch to be considered in the analysis. 
Reasonable intensity and area limits have recently been discussed \citep{moseley2018statistical}. 
All rain cells are tracked using the Iterative Rain Cell Tracking (IRT) method \citep{moseley2014}. {\it Rain tracks} are formed by identifying rain cells that spatially overlap from one time step to the next. 
Each rain track receives a unique identifier, called the {\it track ID} ({\it Details:} \citet{moseley2014}).
Merging and splitting events are handled through a unit threshold ratio, by which the largest cell is continued by maintaining its ID
({\it Details:} \citet{moseley2018statistical}).
 
\noindent
{\bf Tracking CP gust fronts and identifying collisions.}\\
We emit particles within the lowest model level at the boundary pixels of the identified rain cells and advect these particles radially with the horizontal flow. 
The particles are associated with the respective rain event by assigning them a matching track ID. ({\it Details:} Sec.~S2). 

By collocating particles corresponding to different CPs, {\it collisions} between CPs can be mapped out systematically ({\it compare}: Fig.~S4c). 
The collocation is done by scanning a window of $n_{g}\times n_{g}$ grid boxes for the simultaneous occurrence of tracers belonging to distinct CPs ({\it compare:} Fig.~S4), in Fig.~\ref{fig:vertical_velocity}d the strictest and simplest case is presented ($n_{g}=1$), i.e. using only a single grid box for each tracer. 
Examples for $n_{g}=3$ are shown in Fig.~S6.
In cases where more than a threshold number of particles (set to ten in Fig.~\ref{fig:vertical_velocity}d) are identified, the diversity of CP indices within the window is quantified --- leading to the classifications into classes of either zero, one, two, or three CPs (Fig.~\ref{fig:vertical_velocity}).
The sensitivity of our results to $n_{g}$, the threshold number of particles, and the time of the model day were investigated, showing robustness regarding the ordering of the interaction modes (Fig.~S6). In particular, as the day progresses, the distribution function corresponding to 2CP collisions becomes more and more similar to that of 3CP collisions (Fig.~S6f and i). 

\section{Conceptual Model}\label{sec:conceptual_model}
\noindent
Our conceptual model represents CP gust fronts as expanding circles, where the origin of each circle is a given rain event. 
At locations where circles meet, new rain events can form, which then cause new circles to emerge.  \\
\noindent
{\bf Model dynamics.}\\ 
More specifically, $N$ initial cells, referred to as belonging to generation one, are seeded at random locations in a two-dimensional square of area $A\equiv L\times L$, where $L$ is the domain edge length. 
We impose cyclic boundary conditions in both horizontal directions.
The cells grow as circles, representing CP gust fronts. 
All circle radii are initialized as zero and increase linearly in time at constant speed $c_0$.
Our model assumes that CP gust fronts do not travel further after collision. 
Where two circle perimeters eventually collide, an immobile and inactive front is therefore formed. 
Inactive fronts have no further effect.
In each point where three circles collide, a new circle of generation two is seeded with zero initial radius.
As soon as this circle is introduced its radius also grows at speed $c_0$.
The model hence involves a spatial growth process and a replication mechanism.
Besides the restriction to three-CP collisions, our main assumptions are that all CPs spread with constant and equal speed in all directions --- assumptions we revisit in the conclusions.

\noindent
{\bf Model implementation.}\\
The mathematical model ({\it see}: Results) is simulated in two ways: 
(1) a numerical approximate method, where a discrete domain of $2000$ $\times$ $2000$ grid boxes is used, and circle centers are seeded by allocating a unique circle index to $N$ grid boxes selected randomly from a uniform distribution. 
Spreading in space is accomplished, by sequentially copying the circle index to all neighboring grid boxes, which are within the corresponding cold pool radius. 
At locations where grid boxes of three different circle indexes meet and the triangle condition is fulfilled (Fig.~\ref{fig:patterns}a), a new circle center is initialized. 
(2) a semi-analytical method is used as an additional check of the results.
A detailed description of both (1) and (2) is provided in Sec.~S1.


\section{Results}\label{sec:results}
\subsection{Characterizing cold pool interactions}\label{sec:characterizing}
To examine how CPs interact, as a benchmark cold pool numerical experiment, we first analyze LES of the transient response of the convective cloud field to diurnal surface heating ({\it Simulation details:} Sec.~\ref{sec:data}).
Positive near-surface vertical velocity $w(z=100\;m)$, or equivalently, low-level horizontal convergence, is a useful indicator of CP boundaries.
Before the onset of precipitation, the thin boundaries of positive $w(z=100\;m)$ show relatively long-lived cells of typical scales $\sim 2\;km$ \citep[Fig.~\ref{fig:vertical_velocity}a,][]{haerter2017precipitation}.
The geometry of the pattern can be characterized by a network of links (line structures) and nodes (intersects of links) \citep{glassmeier2017network}. 
Both links and nodes, which can result from the interference of two and three CPs, respectively, correspond to updraft regions. 
We want to characterize the nature of the interactions and therefore investigate updraft strength for links versus nodes.
In order to set off new precipitating convection, the updraft speed aloft, near the lifting condensation level ($z\sim 1000\;m$), is relevant. 
High percentiles of $w(1000\;m)$ serve as a proxy for locations where cloud-base velocity is largest (Fig.~\ref{fig:vertical_velocity}a---c).
Visual inspection shows that it is initially nearly always at nodes (quantified here by $w(100\;m)$), rather than along the links of low-level convergence, where $w(1000\;m)$ is largest (Fig.~\ref{fig:vertical_velocity}a), in qualitative agreement with theory on hexagonal Rayleigh-B\'enard cells \citep{drazin2004hydrodynamic}. 

When precipitation sets in at a given location, the gust fronts initially protrude radially, forming near-circular boundaries. 
For a system of many CPs, the radial spreading of a given gust front is eventually reduced or halted by other fronts spreading in opposing directions --- collisions result, again forming boundaries between fronts (new links).
As the pattern evolves, scales increase approximately linearly under the action of CPs \citep{haerter2017precipitation}, qualitatively seen in the coarser convergence patterns (Fig.~\ref{fig:vertical_velocity}b,c). 
Many locations of strong $w(1000\;m)$ again coincide with nodes \citep{boing2012influence}. 
Occasionally links and more rarely single gust front processes also cause new cells to emerge.

To quantify interaction modes, 
tracer particles are seeded near the edges of precipitation cells. The tracer particles are advected with the flow, and thereby highlight the gust fronts belonging to each CP ({\it Details:} Sec.~\ref{sec:methodology} and Fig.~S4).
The histograms of $w(1000\;m)$, conditional on the collocation of particles belonging to different CPs (Fig.~\ref{fig:vertical_velocity}d), quantify the dependence of updrafts on the interaction mode. 
Gust fronts generally are associated with higher vertical velocities.
However, where collisions between gust fronts occur, vertical velocities are larger.
In the case where three gust fronts collide, velocities are even larger, with moderate dependence on the time of the model day (Fig.~S4d---f).
Collisions between two CPs (2CP collisions) conceptually differ from collisions between three CPs: 
In 2CP collisions, air can escape both vertically and tangentially \citep{droegemeier1985three}. 
For 3CP collisions, however, air is horizontally captured in between the three participating CPs. 
Only the vertical degree of freedom remains, leading to exclusively vertical motion. 

\subsection{Applying the conceptual model}\label{sec:application}
We now turn to our conceptual model (Sec.~\ref{sec:conceptual_model}), and describe the dynamics of the diurnal cycle (Fig.~\ref{fig:vertical_velocity}a---c), focusing on the three-CP (3CP) collisions, where strong updrafts are favored.
We first present an analytical mean field solution, which neglects spatial correlations, and subsequently simulate our conceptual model to obtain explicit results on spatial self-organization.

\noindent
{\bf Mean field solution to the conceptual model.}
In a simple approximation, CPs are taken to be seeded simultaneously at discrete iteration steps $n$.
Further, we assume that the cells initiated at each iteration are distributed randomly in space, that is, systematic spatial fluctuations in cell number density are ignored.
Under these assumptions, we estimate the mean distance, $l$, a CP needs to spread before colliding with another as the square root of the CP's mean area. 
With the domain edge length $L$ and the total number of CPs $N$, $l\approx a^{1/2}=L\;N^{-1/2}$, where $a\equiv A\;N^{-1}$ is the mean area available to any individual CP.
The typical time from initiation to collision of a given CP will hence be 
\begin{equation}
\Delta t=L\;N^{-1/2}(2\;c_0)^{-1} \;,
\label{eq:delta_T}
\end{equation}
where the relative speed equals $2\;c_0$, warranting the fact that the CPs approach one-another.

A fraction of the collisions between three CPs is assumed to lead to a new precipitation event.
It is hence meaningful to define a replication rate $r$ as the average number of new cells generated by each present cell. 
To give an upper bound on $r$, we use that in Voronoi graphs for random points, the average number of Voronoi nodes for each point is six \citep{yan2011computing}, hence, six corners are expected on average for each cell.
Given that three cells are involved in each collision, there can be no more than two new CPs for each previous CP, implying $r\leq 2$ ({\it compare:} Figs \ref{fig:vertical_velocity} and \ref{fig:patterns}a).

More explicitly, assuming that the $N(n)$ cells at iteration $n$ produce $N(n+1)=r\;N(n)$ new cells, then the change in the cell number from iteration $n$ to $n+1$ is
\begin{equation}
 \Delta N = (r-1)N(n)\;.
 \label{eq:delta_N}
\end{equation}
To obtain temporal dynamics, using Eqs~\ref{eq:delta_T} and \ref{eq:delta_N}, we can approximate the time derivative of $N(t)$ as 
\begin{eqnarray}
 \frac{dN}{dt}\approx\frac{\Delta N}{\Delta t}=2\;c_0L^{-1}(r-1)N^{3/2}\;,
 \label{eq:dN_dt}
\end{eqnarray}
with $\Delta t$ representing the typical time from one iteration to the next.
Integrating Eq.~\ref{eq:dN_dt} yields 
\begin{equation}
 l(t)\equiv L\;N(t)^{-1/2}=l_0+c_0\;(1-r)t\;,
 \label{eq:l_vs_t}
\end{equation}
where $l(t)$ is the average distance between cell centers at time $t$, and $l_0$ is the integration constant. 
$l_0$ represents the typical spacing between initially seeded cells, hence $l_0=L\;N(0)^{-1/2}$. 
For a replication rate $r>1$ scales shrink, whereas for $r<1$ cells become increasingly sparsely positioned
and scales grow at a constant rate. 
We test this result using previous analysis of spatial scales in LES data \citep{haerter2017precipitation}.
For various different simulations, a linear increase of scale was indeed found (Fig.~\ref{fig:results}a).
Also the slope of the LES simulated curves is reproduced by our analytical model (Eq.~\ref{eq:l_vs_t}) when choosing $r\in [.8,.92]$.
However, $r$ remains an adjustable parameter. 
We therefore now aim to constrain $r$, by demanding that it emerge from the self-organization of the interacting cells.

\noindent
{\bf Self-organized replication dynamics.} 
In our mean field description, we assumed replication to occur at an adjustable rate $r$. 
We now allow $r$ to emerge from physically meaningful constraints on the dynamics.
To this end, consider that not all interferences of three CPs are likely to trigger updrafts:
when three CPs meet, depending on the relative timing and location at which the three originate, two qualitatively different geometries are possible: 
(i) air is enclosed by the three CPs; or 
(ii) air escapes laterally as momentum is channeled away from the location of interference (Fig.~\ref{fig:patterns}a).
The former case can be seen as forcing air masses upward, facilitating updrafts whereas the latter case limits the thrust of the collision.
Case (i) requires the location of collision to lie within the triangle formed by the three circle centers.

To take this distinction into account, we refine our conceptual model by allowing only cases of type (i) and then simulate an initial population of thousands of cells ({\it Details}: Sec.~\ref{sec:conceptual_model}), which all are seeded at time $t=0$ ({\it example}: Fig.~\ref{fig:patterns}b).
The evolving pattern of spreading circles, representing the gust fronts (Fig.~\ref{fig:patterns}c---f), shows a patchwork of mostly small-scale closed circular and larger-scale line structures, with overall length scales increasing over time.
The circular structures at small scales are due to active CP gust fronts that recently emerged and have not yet collided. 
The larger the CP gust fronts become, the more likely they are to have collided with other CPs. 
The stationary fronts resulting from these collisions lead to generally larger-scale line structures. 
The patterns and the scale increase are visually comparable to the self-organizing CP gust fronts in LES (Fig.~\ref{fig:vertical_velocity}).
Also there, it is common to find small circular and larger, more line-like, structures.

\noindent
{\bf Generations.}
Even though any cell of generation $g$ is generally produced at a different time, the positioning of the cell centers corresponding to a given generation nonetheless only allows collisions with cells of the same generation (Fig.~\ref{fig:patterns}b,d,f).
The reason is that the cell center of any given cell of generation $g$ will be enclosed on all sides by cells of generation $g-1$, which themselves expand out from the cell center.
Elements of the dynamics can best be appreciated in a one-dimensional analog (Fig.~S7).

In contrast to the simplified mean field description, cells belonging to any given iteration will not emerge simultaneously. 
However, we use that only cells of equal generations can collide (Fig.~\ref{fig:patterns}b).
We thus assign a label of generation $g=1$ to initial cells and record all subsequent generations: 
any successful collision between generation $g$ cells yields a generation $g+1$ cell.
Collecting all cells of any generation $g$ and computing the average time $t(g)$ when these cells are created, we find that $N(t(g))$ does indeed approximately follow the functional form suggested by Eq.~\ref{eq:l_vs_t} (Fig.~\ref{fig:results}b).
However, the large-$t$ asymptotic behavior shows a statistically significant departure from the exponent $-2$ implied by Eq.~\ref{eq:l_vs_t}.
This deviation requires that the assumption in Eq.~\ref{eq:delta_T} should be generalized as
\begin{equation}
 \Delta t=LN^{-\alpha}(2\;c_0)^{-1}\;,
\label{eq:Delta_t_new}
 \end{equation}
thus yielding a more general version of Eq.~4
\begin{equation}
 N^{-\alpha}=\frac{2\alpha (1-r)}{L}c_0 t+const\;.
 \label{eq:N_alpha}
\end{equation}
In Eqs \ref{eq:Delta_t_new} and \ref{eq:N_alpha}, the exponent $\alpha\neq 1/2$ takes into account that there may be a systematically non-random organization of cells in space. 
In the limit of large $t$, we have $\lim_{t\rightarrow\infty}N(t)\sim t^{-1/\alpha}$, hence, to fit the dependency $\lim_{t\rightarrow\infty}N(t)\sim t^{-1.7}$ (Fig.~\ref{fig:results}b), $\alpha\approx .6$ is required, indicating a measurable departure from a random configuration.

\noindent
{\bf Cell clustering.}
The deviation from $\alpha=1/2$ suggests that cells might be clustered.
To explore this, we analyze the spatial pattern formed by cell centers initiated during a finite time window (Fig.~\ref{fig:results}d and b,inset).
We break the domain down into 400 sub-regions of equal $100\;km\times 100\;km$ areas, and within each sub-region enumerate the number of cell centers during a three-hour time interval. The probability distribution function of all counts in the sub-regions is compared to a shuffled counterpart, where cell centers are distributed at random over the domain during the same time interval ({\it details:} see Sec.~S3).
The analysis shows that the likelihood of finding extreme concentrations of cells in a sub-region increases for the self-organized results relative to random spatial distributions of cells --- a feature that can also be noted qualitatively when inspecting the patchiness of the spatial pattern (Fig.~\ref{fig:results}b, inset).
Repeating for a smaller sub-region (Fig.~S8), on the order of a metropolitan area, we show that clustering occurs also there.

We further point out the non-constant replication rate $r(g)\equiv N(g+1)/N(g)$ (Fig.~\ref{fig:results}c).
For an initially random spatial configuration and synchronous cell initiation, we numerically verify that $r(1)=1$.
The value of $r(g)$ subsequently degrades and eventually saturates near $r\approx .87$ for $n>20$.
Lack of synchrony for $g>1$ indeed is a plausible explanation for this gradual reduction of $r$. 
It is easy to show that two simultaneously seeded cells can ``swallow'' a cell seeded at a later time, that is, the Voronoi cells corresponding to the earlier points can entirely enclose the Voronoi cell corresponding to the later one \citep{kim2001voronoi}.
Such enclosures would reduce the effective number of nodes each point can produce, hence causing $r<1$.

Our model assumes that each 3CP collision yields a new deep convective event, but collisions will likely occasionally fail at doing so, hence $r$ might be somewhat lower than the value obtained here. 
To account for such imperfect replication, appropriately reduced $c_0$ should be considered. 
In reality, however, there may also be other sources of new cells than 3CP collisions.

\section{Discussion} 
Previous work suggested that stronger CPs could enhance precipitation extremes \citep{haerter2018intensified}.
Describing extremes may be accomplished by our model, by coupling precipitation intensity and gust front speed, thereby relaxing the assumption of constant speed $c_0$. 
The actual speed of CP gust fronts, even in a crude approximation, will depend on the initial virtual potential temperature perturbation $\Delta\Theta$ caused by precipitation evaporation, sometimes approximated as $c(\Delta \Theta,h)=\sqrt{2gh(\Delta\Theta/\Theta)}$ \citep{etling2008theoretische}. 
Here, $h$ denotes the effective CP height and $g$ gravitational acceleration. 
We speculate that, when gust fronts of larger speed collide, more environmental air might be enclosed in the resulting updraft. 
Further, due to the effect of surface heat fluxes and energy conservation, the speed of spreading should decrease with radius \citep{gentine2016role,grant2016cold,grant2018cold,romps2016sizes}. The exact dependence on radius hinges on factors such as drag and mixing, which is a matter of current debate.
Extensions of our model that incorporate precipitation intensity and its effect on CP momentum could be used to guide the analysis of extreme precipitation events and how they relate to CP collisions.

Furthermore, our model implies the existence of generations of cells and the lack of interaction between cells of distinct generations. 
It should be considered whether observations, such as those from satellite data, fully support this claim. 
When the gust front of one generation is fast enough to catch up with that of a previous generation, our postulate may be violated. 
In its defense, one should then however also consider the appreciable time delay between an updraft, leading to precipitation, and CP formation. 
Such time delays would diminish the ability of a subsequent gust front catching up with a previous one. 


In reality, CPs cannot spread indefinitely. 
Observed maximal radii in fact vary by an order of magnitude, between $10$ and $100\;km$ \citep{black1978mesoscale,zuidema2012trade,feng2015mechanisms}. 
For larger-scale phenomena, such as the near-planetary organization observed in the Madden-Julian Oscillation \citep{zhang2005madden}, it may be crucial to consider the constraints on maximum CP radius. 
This may also help to better assess models of self-organization, such as those employed in the idealized {\it self-aggregation} case \citep{wing2016self,holloway2017observing,holloway2017convective,wing2017convective,yang2018boundary}.
The detailed effects of CPs on self-aggregation are currently not conclusively understood, but CP strength has been mentioned as modifying the ability of simulations to show self-aggregation \citep{jeevanjee2013convective,muller2015favors}.

Our results on clustering (Fig.~\ref{fig:results}d) may have implications for extreme events, such as flash floods. 
The scales chosen in our example reflect realistic spatial ($\sim 10$---$100$ $km$) and temporal ($\sim 1$---$3\;h$) accumulation intervals \citep{jones2014objective,kendon2014heavier,golding2016mogreps,olsson2017distance} and our analysis suggests that self-organization boosts the likelihood of flash floods compared to random spatial organization.
While qualitative, this finding should be explored in more detail to decipher, which impact clustering can have on the prediction of extremes, today and in a climate with modified near-surface temperature.

Modifications to our model could further allow for refractory effects of CPs:  
the likelihood of updrafts forming in regions previously affected by CPs is reduced due to negative buoyancy resulting from the evaporative cooling. 
Such areas would effectively function as traps, where collisions would not have any pronounced effect in generating updrafts.
Our analysis has further assumed zero-wind shear conditions. 
Wind shear is however known to induce organization along quasi-linear geometries, such as in the case of squall line convection \citep{rotunno1988theory}. 
Variations of our model, which allow for anisotropic spreading of CPs, mimicking the effect of wind shear in squall lines, along with an implementation of negative buoyancy, could be sufficient to allow for the formation of more linear structures as found in squall lines.

As new cells are often generated when three CPs meet, we here developed a conceptual model based on 3CP collisions only. Alternatively, one could imagine conceptual models based on 2CP collisions, or combinations of 2 and 3CP collisions. Such models could also lead to scale changes with time. However, additional considerations, such as the probability of a collision in fact leading to the formation of a new cell center, would have to be taken into account.
Ultimately, generalizations of the model described here could kick-start new attempts at parameterizing convective organization in large scale models.

\section{Conclusion}\label{sec:conclusion}
CP interactions are observed ubiquitously over land \citep{engerer2008surface} and sea \citep{feng2015mechanisms,de2017cold} and large eddy simulations (LES) allow for the analysis of their dynamics \citep{tompkins2001organizationCold,torri2015mechanisms,schlemmer2015modifications,moseley2016,haerter2018intensified}. 
LES incorporate many of the physical processes that are relevant in describing the formation and dynamics of CPs: three-dimensional fluid dynamics with appropriate boundary conditions, condensation and evaporation effects, precipitation formation, as well as radiative transfer. 
Using LES, we here pinpoint in which ways CPs interact during the diurnal cycle, and that the collision of gust fronts from three CPs is crucial for cloud formation.
We then formulate a simple conceptual model where we represent CPs as circles that grow with equal and constant radial speed. The model is initialized with circles that expand from randomly located points in 2D space. Replication of CPs is introduced by seeding a new expanding circle from the point where three circles collide. We show that this model allows us to reproduce the dynamics of CP gust fronts found in the LES.
Our model, intentionally left simple, captures the essential aspects of diurnal cycle dynamics, where spatial scales increase linearly with time. 
The model highlights the complexity of convective self-organization through CPs, and offers a framework for the description of clustering and extremes. 

\pagebreak
\section*{Acknowledgments}
We thank Guillermo Garc\'ia-Per\'ez, Christopher Moseley, Kim Sneppen and Peter Berg for fruitful discussions on the mathematical model.
We are grateful to both anonymous reviewers for their in-depth comments to our manuscript.
OH and JOH gratefully acknowledge funding by a grant (13168) from the VILLUM Foundation.
This project has received funding from the European Research Council (ERC) under the European Union's Horizon 2020 research and innovation program (grant agreement no. 771859). 
SJB is partially funded through the NERC/Met Office Joint Programme on Understanding and Representing Atmospheric Convection across Scales (GENESIS, grant number NE/N013840/1).
SBN acknowledges funding through the Danish National Research Foundation (grant number: DNRF116). 
The simulation data and program code used for the LES analysis are available from \citet{moseley2016}.

\pagebreak
 \newcommand{\noop}[1]{}


\pagebreak
\begin{figure}
\begin{center}
\caption{{\bf Analysis of cold pool collisions.} 
{\bf a---c},
Horizontal cross sections of large-eddy simulated near-surface vertical velocity $w(100\;m)$.
Darker shades qualitatively highlight larger vertical velocity (quantification: {\it see} Fig.~S5).
Blue symbols denote locations with $w(1000\;m)$ exceeding the $99.98$ $th$ percentile. 
Panels correspond to times near precipitation onset ($11$ $h$), the time near greatest domain mean intensity ($15$ $h$) and for decaying mean intensity ($18$ $h$), respectively.
({\it Details}: Fig.~S3).
Note, the axis scales are identical in all three panels, highlighting a progressive scale increase.
{\bf d}, Probability density functions (PDFs) conditional on the geometric location within CPs, the legend labels the number of CPs contributing, for all times where precipitation was present in the LES (all $t$ where $I(t)>0$ in Fig.~S3). 
Short vertical lines near the horizontal axis show means of corresponding PDFs. ({\it Technical details:} Sec.~\ref{sec:methodology}).
}
\label{fig:vertical_velocity}
\end{center}
\end{figure}

\pagebreak
\begin{figure*}
\begin{center}
\caption{{\bf CP dynamics in the conceptual model.}
{\bf a}, Increase of spatial scales for LES for various boundary conditions ({\it Details:} Sec.~\ref{sec:data}).
$t_0$ marks the time of precipitation onset for each simulation.
The speed of scale increase $v\equiv c_0(1-r)$ is the slope of a fit to the data points (exemplified for P4K, black line, where $v\approx .8\;m\;s^{-1}$) \citep{haerter2017precipitation}. 
Using $v=.8\;m\;s^{-1}$, Eq.~\ref{eq:l_vs_t} and $c_0\in[4,10]\;m\;s^{-1}$, estimated from LES, the range, blue arrow in (c), is obtained.
Plot modified from \citet{haerter2017precipitation}.
{\bf b}, Number of cells per generation $N(g)$ plotted against the corresponding average creation time $t(g)\equiv N(g)^{-1}\sum_i t_i^{(g)}$: each point represents the number of cells in a given generation.
Time is measured in units of $L/c_0$, arrows along panel top show approximate time scales for typical values of $c_0$ and initial cell density.
The black and green straight dashed lines represent power laws with exponents $-1.7$ and $-2$, respectively.
Labels along the curve indicate several generation numbers for clarity.
{\it Inset:} conceptual model results in a $800\times 800$ $km^2$ sub-area after approximately $3h$ from initial random seeding of cells, during a $3h$ interval. 
We assume an initial average cell-to-cell distance of $10\;km$ and $c_0\approx 10\;ms^{-1}$. ({\it Details:} Sec. S3).
{\bf c}, Replication rate $r(g)$ from 3CP model (black curve) and estimate from LES (blue arrow), using (a).
{\bf d}, Cumulative distribution functions for simulated (blue) and randomized (red) cell count for data as in the inset to (b). Cell count is normalized to the mean cell count in the sub-domain ({\it Details:} Sec.~\ref{sec:conceptual_model}).
Note, the logarithmic vertical axes in (b) and (d) and the logarithmic horizontal axes in (b) and (c). 
}
\label{fig:results}
\end{center}
\end{figure*}

\begin{figure*}[b]
\begin{center}
\vspace{30pt}
\caption{{\bf Model and convective scale increase.} 
{\bf a}, Schematic illustrating three-CP collisions: for successful collision, the collision point (bold black dot) must lie within the triangle formed by the three circle centers ($\mathbf{c}_1$,$\mathbf{c}_2$,$\mathbf{c}_3$) (thin black dots) shown. The red line illustrates the stationary gust front.
{\bf b}, Example of CP spreading in space (different colors indicate different CPs). 
Panels from left to right, first and second row, indicate increasing time. 
Note the formation of stationary fronts when CPs interfere. 
Upper right panel: note the junction of three CPs without formation of a new CP (marked by a black arrow); 
lower panels: note the formation of several new CPs (one is highlighted by a green arrow). 
Further note the formation of curved fronts between these new CPs due to asynchronous timings in the second and further generations (white arrow).
{\bf c}, Gust fronts simulated by the three-CP model approximately $2.5\;h$ after precipitation onset when using initial cell density $N(0)A^{-1}=.01\;km^{-2}$ and $c_0=10\;m\;s^{-1}$. 
Lighter gust front shades mark older, stationary, fronts.
{\bf d}, Cell generations corresponding to the areas covered by the different CPs in (c).
White number mark cold pool generation corresponding to color shading, i.e., bright red corresponds to latest generation.
{\bf e,f}, Similar to (c) and (d), respectively, but after approximately $5\;h$ after precipitation onset. 
Note the temporally increasing spatial scale, when comparing (c) and (e), or (d) and (f), respectively.
}
\label{fig:patterns}
\end{center}
\end{figure*}

\pagebreak
\newpage
\clearpage
\begin{figure*}
\begin{center}
\includegraphics[width=12cm]{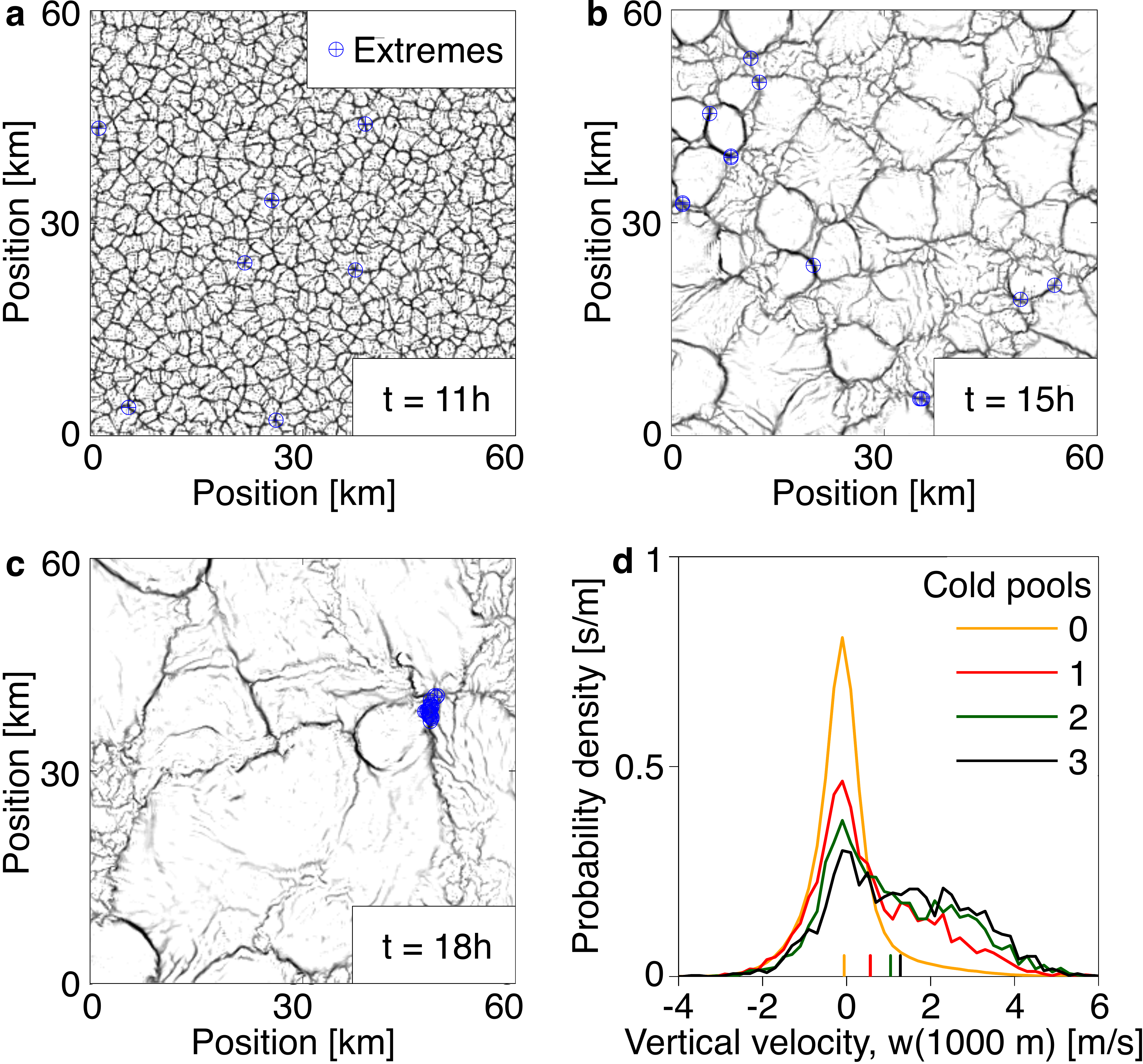}
\end{center}
\end{figure*}

\pagebreak
\newpage
\clearpage
\begin{figure*}
\begin{center}
\includegraphics[width=12cm]{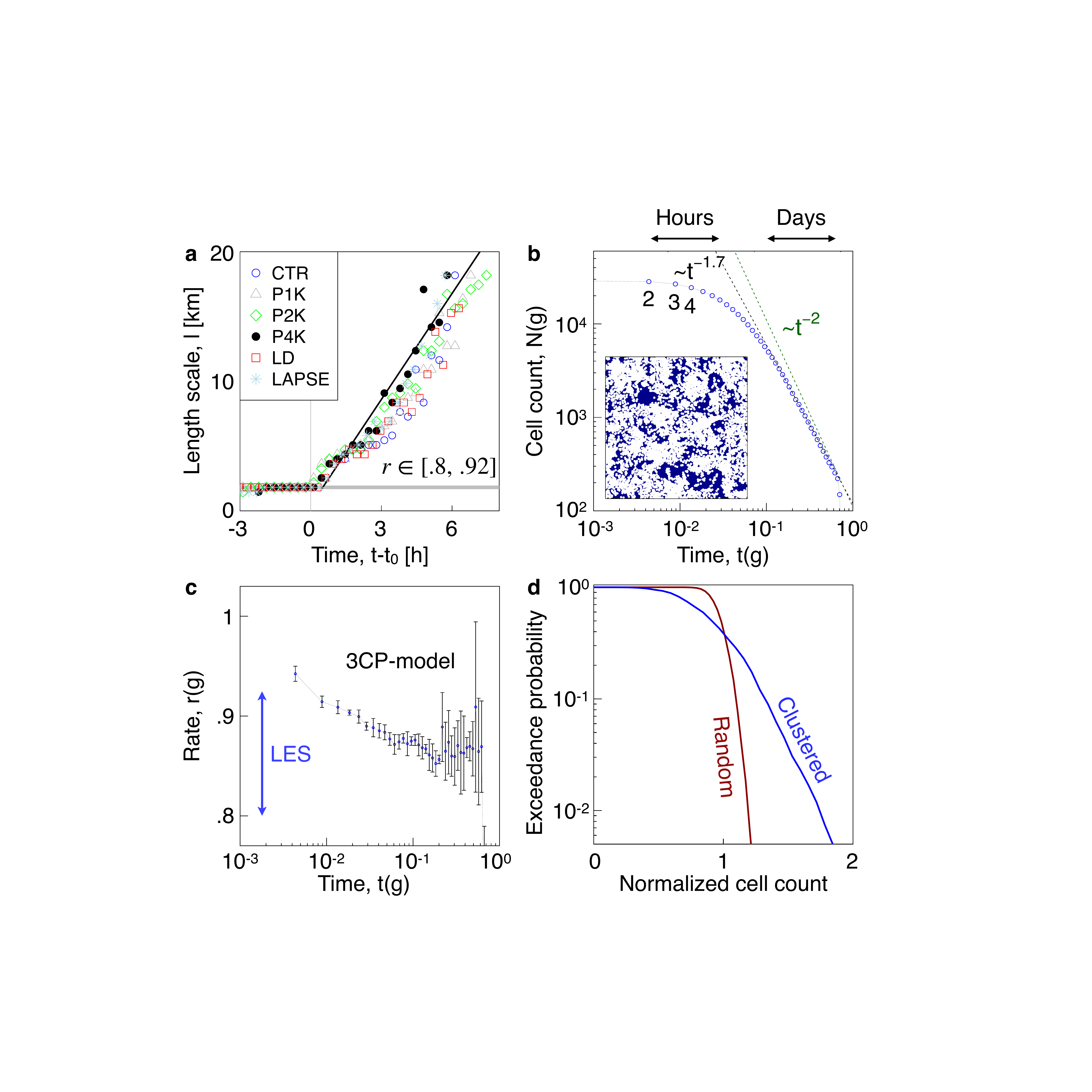}
\end{center}
\end{figure*}

\pagebreak
\newpage
\clearpage
\begin{figure}[b]
\begin{center}
\includegraphics[width=16cm]{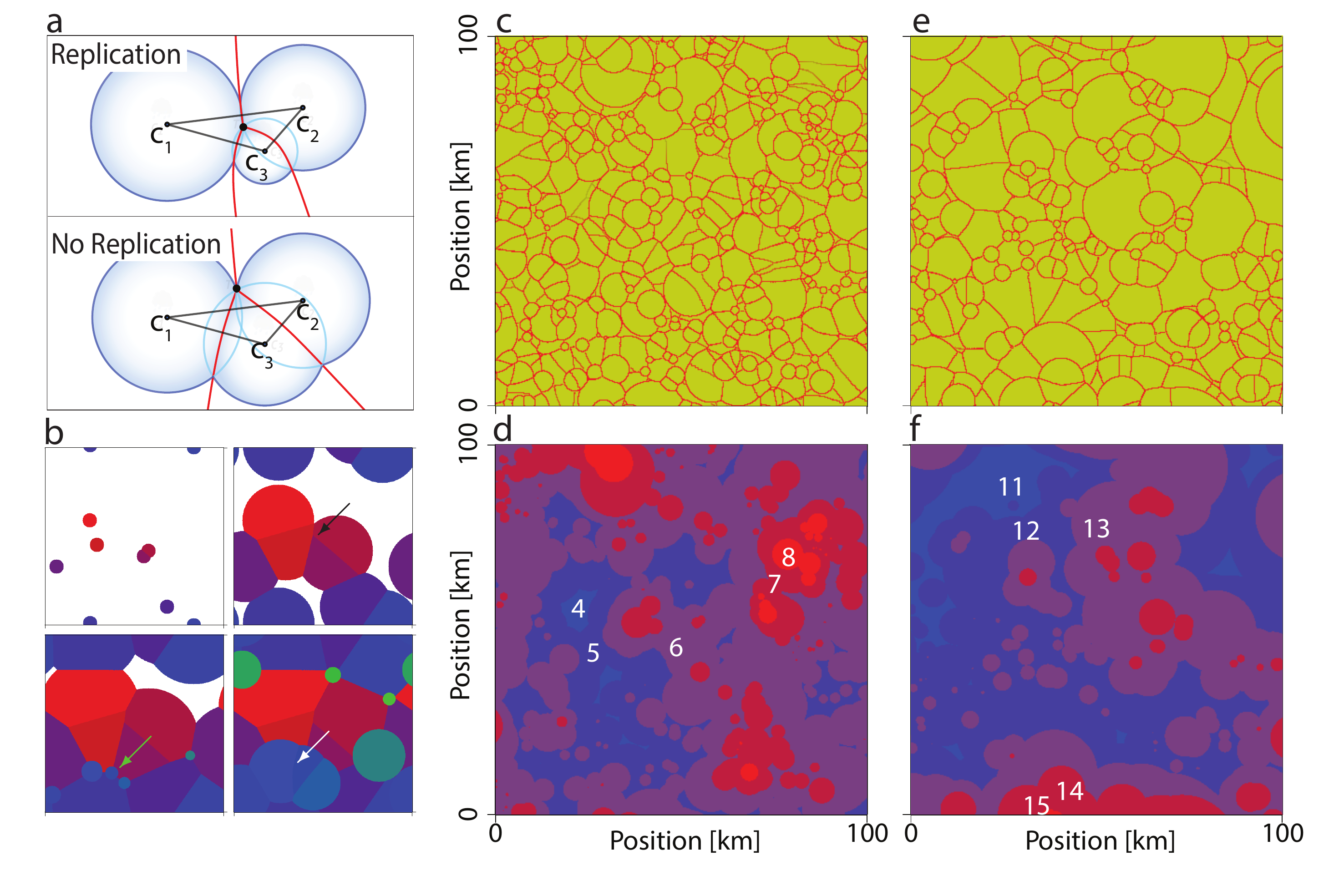}
\end{center}
\end{figure}

\end{document}